\documentclass[twocolumn,showpacs,pra,aps]{revtex4}

\usepackage{graphicx}
\usepackage{dcolumn}
\usepackage{bm}

\begin{document}

\title{Length Distribution of Periodic Orbits \\ of
Unitary Discrete Tent Maps}

\author{Yuriy E. Kuzovlev}
\email{kuzovlev@kinetic.ac.donetsk.ua}
\affiliation{A.A.Galkin Physics and Technology Institute
of NASU, 83114 Donetsk, Ukraine}

\date{\today}

\begin{abstract}
The discrete unitary (reversible) analogues of the continuous
(irreversible) tent maps are numerically investigated, in
particular, the lengths probability distribution of their periodic
orbits. It is found that its density can be well approximated by
the inverse proportional law.
\end{abstract}

\pacs{05.45.-a}

\maketitle

\section{Introduction}
As known, many key features of continuous-time chaotic dynamics
can be treated in terms of two-dimensional or even one-dimensional
discrete-time maps (see e.g. \cite{ll}). Among them, most simple
and perfect ones are the famous Bernoulli map,
 $\,X_{n+1}=$ $2X_n\,$mod$\,1\,$, and the tent map,
 $\,X_{n+1}=$ $1-|1-2X_n|\,$, both defined at the interval
 $\,0\leq X\leq 1$. Almost any trajectory of both maps is chaotic,
that is dense in all this interval.

In computer, however, any trajectory finishes at zero after
$\,\leq D\,$ steps only, where $\,D\,$ is number of bits under
use. This contraction of the discretized phase space, from total
$\,N=2^D\,$ points to single point $X=0$, happens because the maps
are non-invertible, i.e. relation between $\,X_n\,$ and
$\,X_{n+1}\,$ is not an one-to-one correspondence. But a suitable
slight distortion can make a discretized map invertible.
Evidently, any such map is nothing but a permutation of
$\,N=2^D\,$ states ($\,X_n\in$ $[0:N-1]/N$).

In particular, the invertible discrete versions of the Bernoulli
map describe the Mersenne digital auto-generators of pseudo-random
0-1-sequences (see e.g. \cite{dix}). For example, the Mersenne map
in Fig.1 represents transitions between 16 states of a particular
``toy'' 4-bit Mersenne generator ($D=4$). The general formula of
such maps is
\begin{equation}
\begin{array}{lcr}
  Y=2X+1-\sigma(X) & , & \,X<N/2\,\,;\\
  Y=2X-N+\sigma(2X-N) & , & \,X\geq N/2\,\,,
\end{array}  \label{dbm}
\end{equation}
where $\,Y\equiv NX_{n+1}\,$, $X\equiv NX_n\,$, and any of
$\,\sigma(k)\,$, $\,k=0,1,..,N/2-1\,$, equals to either zero or
unit, $\,\sigma(k)=0,1$ (hence this is class of $\,2^{N/2}$
different maps). To see corresponding unitary evolution matrix, it
is necessary to turn the map picture headfirst and insert zeros
and ones in place of the empty cells and crosses, respectively.

The map in Fig.1 has two immovable points ($X=0$ and $X=15$) and
two cycles (periodic orbits) with equal lengths $\,L=7$. In
applications of much greater Mersenne maps (with $D\sim 70\,$
\cite{dix}), the main issue is construction of those having most
long cycles, with $\,L\sim N$.

Here, we will be interested in modest problem about statistics of
cycles of discrete maps, to be concrete, definite class of
discrete tent maps.
\begin{figure}
\includegraphics{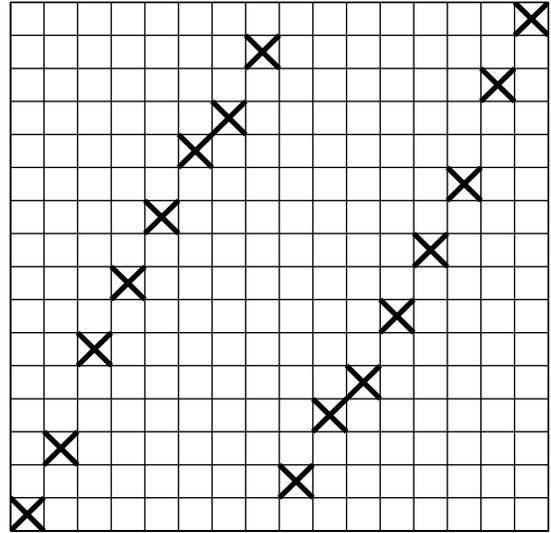}
\caption{\label{fig1} Example of the discrete Bernoulli map (the
Mersenne map, see body text).}
\end{figure}

\section{Cycles of continuous tent maps}
\begin{figure*}
\includegraphics{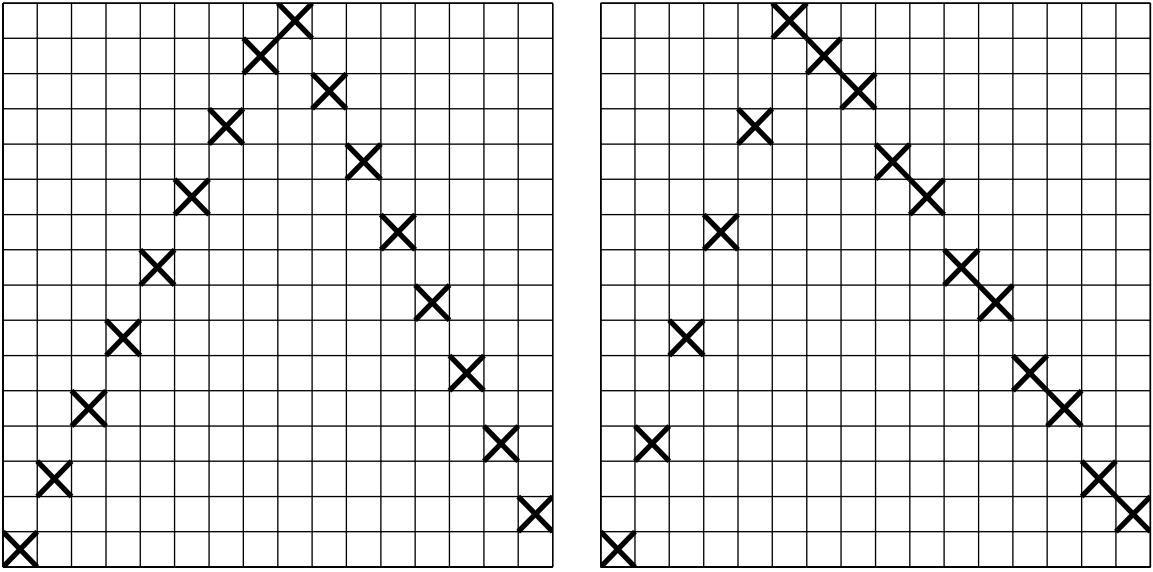}
\caption{\label{fig2}Two examples of discrete tent maps at
$\,D=4\,$ and $\,a=1/2\,$ (left) and $\,a=1/3\,$ (right).}
\end{figure*}
First consider the variety of generally asymmetric continuous tent
maps (CTM) described by
\begin{equation}
\begin{array}{lcr}
  X_{n+1}=X_n/a & , & \,\,X\leq a\,\,;\\
  X_{n+1}=(1-X_n)/(1-a) & , & \,X>a\,\,, \label{ctm}
\end{array}
\end{equation}
where $\,0<a<1$. The inversion of (\ref{ctm}) reads as
\begin{equation}
  X_{n-1}=S_n+(a-S_n)X_n\,\,,\,\,\,
  S_n=0\,\,\text{or}\,\,1\,\,,\label{ictm}
\end{equation}
with $\,S_n=0$ and $\,S_n=1$ representing two variants of the
inversion. Let us write, with the help of (\ref{ictm}), $\,X_0\,$
as a function of $\,X_L\,$ and then equate $\,X_0\,$ to $\,X_L\,$
at various possible values $\,S_1,..,S_L$ $=0,1\,$. Thus we obtain
$\,2^L\,$ linear equations which determine $\,2^L\,$ points
belonging to all possible periodic orbits (cycles) of length
$\,L\,$. That are either irreducible (indivisible) orbits or, if
$\,L\,$ has a divider $\,l\,$, composite orbits consisting of
$\,L/l\,$ repetitions of an irreducible orbit with length $\,l\,$.
Consequently,
\begin{equation}
\sum_{
\begin{array}{c}
  1\leq l\leq L\\
  L\,\text{mod}\,\,l\,=0
\end{array}
}\,l\,N_l\,=\,2^L\,\,\,,\label{expan}
\end{equation}
where $\,N_l\,$ is number of different irreducible periodic orbits
with length $\,l\,$, and the sum is taken over all the dividers of
$\,L\,$.

In particular, if $\,L\,$ is a prime number then (\ref{expan})
yields
\begin{equation}
N_l=(2^L-2)/L\,\,\,(L\,\,\text{is prime})\,\,,\label{prime}
\end{equation}
since $\,N_1=2$ (there are two immovable points). In general, it
is not hard to derive from (\ref{expan}) that
\begin{equation}
2^L-2\left(2^{[L/2]}-[L/2]\right)\leq LN_L\leq
2^L-2\,\,\label{any}
\end{equation}
Here $\,[L/2]$ $=L/2$ if $\,L$ is even, and $\,[L/2]$ $=(L-1)/2$
if $\,L$ is odd. At any $\,L\gg 1$, according to (\ref{any}),
\begin{equation}
N_L \approx 2^L/L\,\,\label{asympt}
\end{equation}
At relatively small $\,L$, from (\ref{expan}) one finds:
$\,N_2=1$, $\,N_3=2$, $\,N_4=3$, $\,N_5=6$, $\,N_6=9$, $\,N_7=18$,
$\,N_8=30$, $\,N_9=56$, $\,N_{10}=99$, $\,N_{11}=186$, ...

\section{Discrete tent maps}
Next, consider invertible discrete tent maps (DTM) representing
discrete analogues of the continuous tent maps (\ref{ctm}). To be
concrete, let us introduce the class of discrete maps defined by
\begin{equation}
\begin{array}{lcr}
  Y=\left[\frac {NX}{A}\right] & , & \,X<A\,\,;\\
  Y=\left\{\frac {N(N-X)}{N-A}\right\}-1 & , & \,X\geq A\,\,
\end{array}   \label{dtm}
\end{equation}
Here, like in (\ref{dbm}), $\,X=NX_n$ and $\,Y=NX_{n+1}$ are
integers, $\,0\leq X,Y$ $\leq N-1\,$; $\,A\,$ is analogue of
$\,a\,$ in (\ref{ctm}), $\,0\leq A\leq N-1$; $\,[x]\,$ is an
integer closest to $\,x\,$ from below (with $\,[x]$ $=x\,$ if
$\,x$ is integer), and $\,\{x\}\,$ means closest integer greater
than (or equal to) $\,x\,$ (that is $\,\{x\}$ $=x\,$ if $\,x$ is
integer). Two examples of maps defined by formula (\ref{dtm}), at
$\,D=4$, are shown in Fig.2.

Let us prove that (\ref{dtm}) prescribers invertible maps, i.e.
establishes an one-to-one correspondence between $\,X$ and $\,Y$.
It is sufficient to prove that the upper r.h.s. and lower r.h.s.
in (\ref{dtm}) can not produce one and the same number. Indeed, in
such a case we would have
\[
\left[NX^{\prime}/A\right]=Y=\left\{
N(N-X^{\prime\prime})/(N-A)\right\}-1\,\,,
\]
with some $\,X^{\prime}<A$ and $\,X^{\prime\prime}\geq A$. This
would mean that
\[
\begin{array}{c}
NX^{\prime}=AY+u\,\,,\,\,\,0\leq u<A\,\,,\\
N(N-X^{\prime\prime})=(N-A)Y+v\,\,,\,\,\,0<v\leq N-A\,\,,
\end{array}
\]
($\,u\,$ and $\,v\,$ are some integers), therefore,
\[
\begin{array}{c}
N+X^{\prime}-X^{\prime\prime}=Y+(u+v)/N\,\\
0<u+v<N
\end{array}
\]
But, obviously, the latter inequality contradicts the preceding
equality. The proof is finished.

\begin{figure*}
\includegraphics{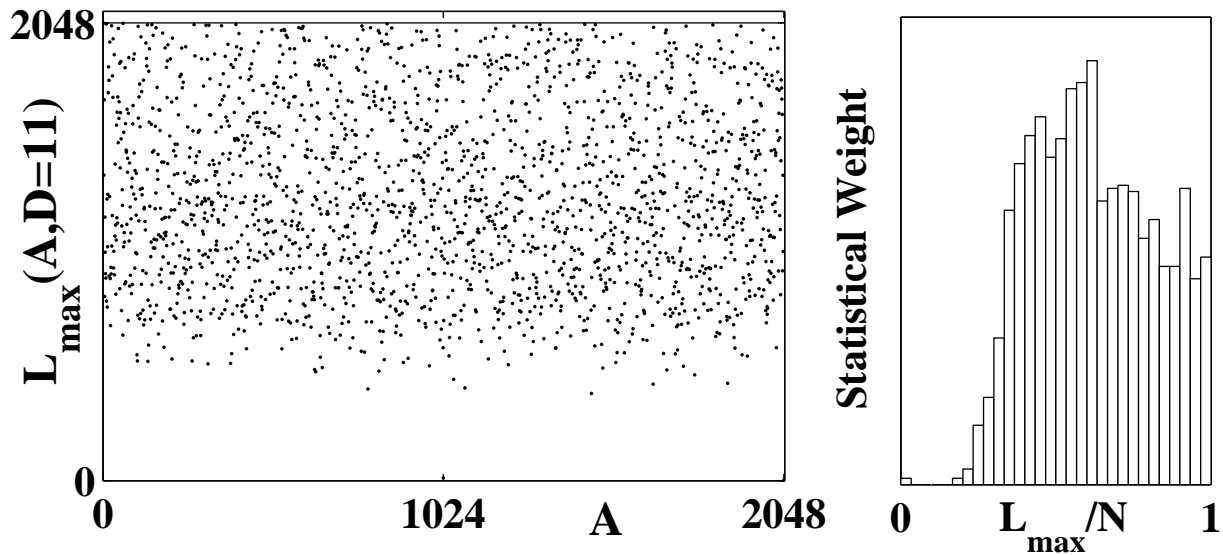}
\caption{\label{fig3}\,\,(On the left)\, The length,
$\,\,L_{max}\,$, of most long periodic orbit of the discrete tent
map, as a function of the asymmetry parameter of the map, $\,A\,$,
at $\,N=2048$ ($\,D=11$). \,(On the right) The corresponding
histogram of $\,L_{max}/N$.}
\end{figure*}
It is easy to see also, that absolute value of a deviation of any
DTM (\ref{dtm}) from its CTM prototype (\ref{ctm}), with
$\,a=A/N$, (as well as deviation of (\ref{dbm}) from Bernoulli
map) does not exceed $1$ lowest bit only ($\pm 1/N$, in terms of
$\,X_n$). In this sense, any DTM tends to corresponding CTM, when
$\,N\rightarrow\infty$. However, the inverted map remains strongly
discontinuous, which is a payment for its unambiguity.

Hence, in essence, the limit of DTM at $\,N\rightarrow\infty$ does
not coincide with corresponding CTM.

\section{Cycles of discrete tent maps}
Naturally, there is no simple rule for the cycles (periodic
orbits) of the DTMs. A number of various irreducible cycles and
lengths of these cycles are extremely irregular functions of
$\,A\,$ and $\,N$. In contrary to (\ref{expan}), now
\begin{equation}
\sum_{1\leq L\leq N} L\,N_L\,=\,N\,=\,2^D\,\,,\label{expan_}
\end{equation}
where $\,N_L$ is a number of different irreducible periodic orbits
with length $\,L\,$ (possibly, $\,N_L=0$). For example, at
$\,D=12$ and some three next values of $\,A$ this expansion looks
as
\[
\begin{array}{c}
1+4095=N\,\,,\\
1*2+2*2047=N\,\,,\\
1+13*315=N\,,\\
\end{array}
\]
where the second multiplier (if any) in each term is number of
different cycles of a length represented by the first multiplier.
In other example, for $\,D=15$,
\[
\begin{array}{c}
1*2+2+4*3+8*30+16*2032=N\,\,\,(A=N/2)\,,\\
1+32767=N\,\,\,(A=N/2-1)\,,\\
1*2+2+4+8*5+16*65+24*2+32+48*9+\\
+60*2+64+72+103+112*20+120+128*10+\\
+144+176+192*6+224*4+240+1200+\\
+1570+8792+12999=N\,\,\,(A=N/2+1)
\end{array}
\]

Typically, a DTM have a long cycle whose length is comparable with
the total number of points, $\,N$, i.e. maximal possible length.
This is illustrated by Fig.3 relating to $\,D=11$. Quite similar
pictures take place also at greater $\,D$. We see that practically
any DTM has a cycle with length $\,L\gtrsim N/4\div N/3$. Such
long cycles, whose length $\,L$ is comparable with $\,N$, can be
treated as discrete analogues of the chaotic trajectories of CTM.

The left plot in Fig.3 demonstrates highly irregular dependence of
the maximal cycle length, $\,L_{max}$, on the asymmetry parameter
$\,A$. At the same time, the right-hand plot there shows that even
slight smoothing of the data produces rather regular results.
Therefore, it is reasonable to describe the cycles of DTMs in
statistical language, considering some specific subclasses of DTMs
instead of individual maps.

\section{Statistics of cycles of symmetric discrete tent maps}
Let us consider the family of nearly (asymptotically) symmetric
DTMs determined by the conditions
\begin{equation}
\frac {N-M}{2}\leq A<\frac {N+M}{2}\,;\,\,\,
N,M\rightarrow\infty\,; \,\,\, \frac {M}{N}\rightarrow 0
\label{sym}
\end{equation}
Since $\,M\rightarrow\infty $, we obtain the growing statistical
ensemble, which is all the more representative one because all the
maps defined by (\ref{sym}) do tend to the same limit
$\,A/N\rightarrow$ $a=1/2$.

\begin{figure*}
\includegraphics{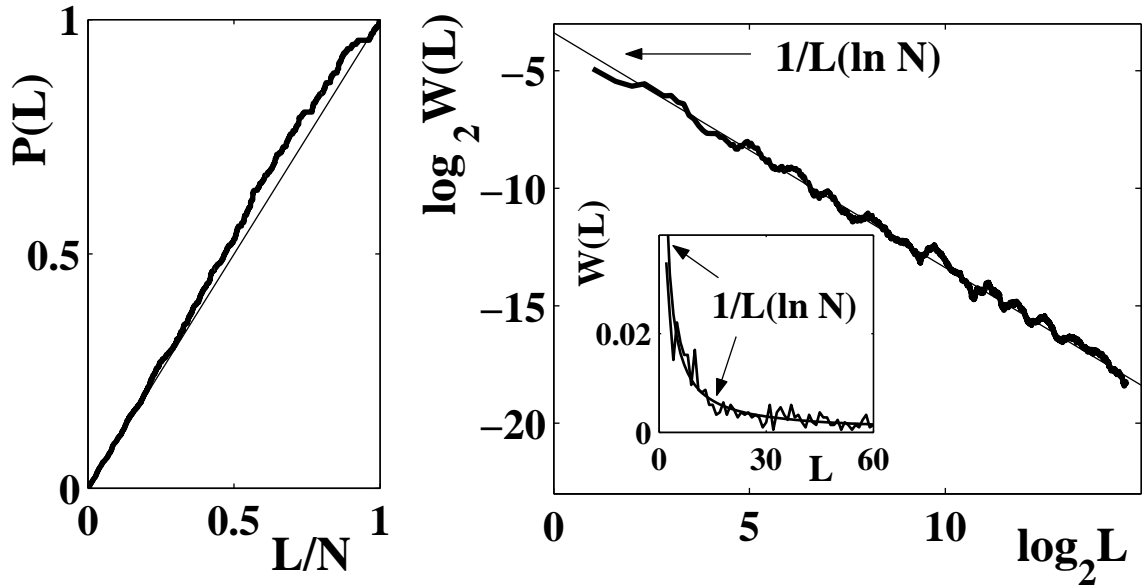}
\caption{\label{fig4}\,(Left plot) The probability, $\,P(L)$, that
an arbitrarily chosen point belongs to some periodic orbit (of the
nearly symmetric discrete tent maps (\ref{sym})) with lengths less
than or equal to $\,L$, as a function of $\,L/N$, at $\,D=14$ and
$\,M\sim 200$. (Right plot) The binary logarithm of the
probability, $\,W(L)\,$, that arbitrarily chosen periodic orbit
(of a similar maps family) has the length $\,L$, via binary
logarithm of $\,L$, at $\,D=15$ and $\,M\sim 500$, in comparison
with the analytical estimate of the $\,W(L)\,$ dependence (the
straight line). The inset demonstrates the same comparison for
small lengths.}
\end{figure*}
It would be interesting to investigate such the family of
asymptotically symmetric discrete tent maps (ASDTM), in comparison
with the usual symmetric CTM. Here, simplest statistical
characteristics of cycles (periodic orbits) of the ASDTM will be
under our attention.

Let $\,W(L)\,$ designates a density of probability distribution of
the cycle lengths in this family of maps. In other words,
practically,
\begin{equation}
W(L)=N_L/\sum_{l=1}^{N}N_l\,\,, \label{W}
\end{equation}
where $\,N_L$ is now {\bf{\it summary}}~ number of periodic orbits
of length $\,L$ in all the maps of the ASDTM family. It is useful
to introduce also the quantity
\begin{equation}
P(L)=\frac {1}{MN}\sum_{l=1}^{L}lN_l\,\,, \label{P}
\end{equation}
that is relative (probability) measure of points which belong to
all periodic orbits with lengths $\leq L\,$ at all the maps.

In reality, with the help of an ordinal PC only, it would take
rather long time to obtain all the periodic orbits if $\,D\geq
16\div 17$. But our computations, performed at $\,D=12\div 16$,
showed that already $\,D=13\div 14$ are satisfactory values,
because next $\,D$'s increases do not change the picture
qualitatively (although, of course, providing better numeric
accuracy).

Therefore, it is reasonable to prefer calculations at not high
$\,D$ ($D=14\div 15$) but apply slight smoothing over $\,L$'s
values. Concretely, the plot $\,W(L)$ in Fig.3 represents the
result of averaging of the exact $\,W(L)$ (defined by (\ref{W}))
over the intervals $\,[L,L+\min(\delta L,1)]\,$, with
$\,\delta\sim 0.01\,$.

Such smoothed probability density, $\,W(L)$, is represented by the
curved line at right-hand side of Fig.3. It is easy to find that
the best fitting for it is nothing but the inverse proportional
law:
\begin{equation}
W(L)\approx \frac {1}{L\,\ln N}\,\,\,, \label{ip}
\end{equation}
where factor $\,(\ln N)^{-1}$ ensures the normalization,
\[
\sum_L W(L)\approx \int_1^N W(L)dL=1
\]
The dependence (\ref{ip}) is shown by the straight line, and by
smooth curve in the inset which demonstrates good quality of this
fitting for short cycles too.

The curved line at left side of Fig.3 shows an example of the
probability (\ref{P}). It is not smoothed, therefore, formed by
many steps with very different heights (like the famous devil's
staircase). Its closeness to the thinner straight line, which
corresponds to $\,P=L/N$, says that the total $\,MN$ points are
distributed approximately equally between cycles with different
lengths. This is just about what the approximation (\ref{ip})
says.

If combining (\ref{W}),(\ref{P}),(\ref{ip}) and the approximation
$\,P=L/N$, we obtain the estimate of mean number of cycles per one
map of the family:
\begin{equation}
\frac {1}{M}\sum_{L=1}^{N}N_L\approx \,\ln\,N\,, \label{mnc}
\end{equation}
(let us recollect that $\,N_L$ is {\bf{\it summary}} value for all
the ASDTM). At the same time, as the above examples of map
expansions into cycles do show, fluctuations in number of cycles
from one particular map to another are significantly greater than
the mean value (\ref{mnc}). According to these examples, as well
to the $\,W(L)\,$ plots, especial contribution to the fluctuations
comes from cycles whose lengths are powers of two.

Nevertheless, when rising $\,D$ from 12 to 16 a definite decrease
of the fluctuations was noticed. This observation pushes us to the
hypothesis that formula (\ref{ip}) represents a true asymptotics
of the cycles length distribution (\ref{W}) in the limit
(\ref{sym}).

\section{Discussion and resume}
There is a simple naive explanation of the hypothetical
asymptotics (\ref{ip}). In above derivation of the estimate
(\ref{asympt}) for the cycles numbers of continuous tent maps
(CTM), the factor $\,2^L\,$ (with $\,L$ being cycle lengths)
arises from the two-valued property of their inverse maps (i.e.
due to their irreversibility). Since the discrete tent maps (DTM)
under consideration have univalent inversions, in their case this
factor must disappear. Thus one deduces the inverse proportional
dependence of number of the cycles (periodic orbits) on their
lengths, i.e. comes to (\ref{ip}) (in other words, cycles of
different lengths involve approximately equal amounts of points of
the discrete phase space).

In view of such reasonings, the inverse proportional law can be
expected in case of any sufficiently reach family of unitary DTM,
not only the ASDTM family defined by (\ref{sym}) (moreover, in
unitary discrete analogues of multi-modal piecewise linear CTM,
not unimodular tent maps only).

In factual finite-precision computer simulations of more or less
general CTM, distinguished from the ``pathological'' map
$\,X_{n+1}=$ $1-|1-2X_n|\,$ (about it see Introduction), e.g.
asymmetrical CTM (\ref{ctm}) or the map $\,X_{n+1}=$
$1-c|1-2X_n|\,$ with e.g. $\,c=0.9999999999..\,$, it can be
expected that the law (\ref{ip}) must manifest itself sooner than
(\ref{asympt}). Indeed, if the latter rule was true, then the very
short orbits with lengths $\,L\leq D\,$ only (recollect that $\,D$
is number of bits under operations) would take all $\,N=2^D$
points of the discrete phase space. In reality, with no doubts,
almost any choice of an initial point results in a long orbit
(``chaotic trajectory''), whose length $\,L$ is comparable with
$\,N$, while it is hard to hit casually into a short orbit.

To resume, we performed computer statistical analysis of periodic
orbits of unitary (reversible) discrete tent maps, and found that
probability distribution of the orbits lengths well obeys the
inverse proportional law (\ref{ip}).

I am grateful to Dr. I.Krasnyuk for useful conversations.

\begin{verbatim}
\end{verbatim}




\end{document}